%% file: main.tex
\documentclass[runningheads]{llncs}

\input{preamble}

\usepackage[T1]{fontenc}

\begin{document}

\title{RefinementEngine: Automating Intent-to-Device Filtering Policy Deployment under Network Constraints}

\titlerunning{RefinementEngine}

\author{Davide Colaiacomo\inst{1}\orcidID{0009-0001-1761-5218} \and
Chiara Bonfanti\inst{1}\orcidID{0009-0007-8015-7786} \and
Cataldo Basile\inst{1}\orcidID{0000-0002-8016-1490}}

%\author{Author name 1\inst{1}\orcidID{0000-0000-0000-0000} \and
%Author name 2\inst{1}\orcidID{0000-0000-0000-0000} \and
%Author name 3\inst{1}\orcidID{0000-0000-0000-0000}}

\institute{Politecnico di Torino, Torino, Italy - \email{\{name.surname\}@polito.it}}
%\institute{University name, city, country - \email{Authors emails}}

\maketitle              

\begin{abstract}
Translating security intent into deployable network enforcement rules and maintaining their effectiveness despite evolving cyber threats remains a largely manual process in most Security Operations Centers (SOCs). In large and heterogeneous networks, this challenge is complicated by topology-dependent reachability constraints and device-specific security control capabilities, making the process slow, error-prone, and a recurring source of misconfigurations.
This paper presents RefinementEngine, an engine that automates the refinement of high-level security intents into low-level, deployment-ready configurations. Given a network topology, devices, and available security controls, along with high-level intents and Cyber Threat Intelligence (CTI) reports, RefinementEngine automatically generates settings that implement the desired intent, counter reported threats, and can be directly deployed on target security controls.
The proposed approach is validated through real-world use cases on packet and web filtering policies derived from actual CTI reports, demonstrating both correctness, practical applicability, and adaptability to new data. 

\end{abstract}

\section{Introduction}
\label{sec:intro}
\input{sections/introduction}

\section{Background}
\label{sec:background}
\input{sections/background}

\section{Implementation}
\label{sec:implementation}
\input{sections/implementation}

\section{Validation}
\label{sec:validation}
\input{sections/validation}

\section{Related works} 
\label{sec:related}
\input{sections/related}

\section{Conclusions}
\label{sec:conclusions}
\input{sections/conclusions}

%\begin{credits}
%\subsubsection{\ackname} ...

%\subsubsection{\discintname} ...
%\end{credits}

\bibliographystyle{splncs04}
\bibliography{bibliography}

\end{document}

%% file: preamble.tex
\usepackage{graphicx}
\usepackage{cite}
\usepackage{url}
\usepackage{float}
\usepackage{listings}
\usepackage{xspace}
\usepackage[acronym, nomain]{glossaries}
\usepackage[colorlinks=true, linkcolor=blue, urlcolor=blue, citecolor=blue]{hyperref}
\usepackage{tikz}
\usepackage{xurl}

\newacronym{soc}{SOC}{Security Operations Center}
\newacronym{hspl}{HSPL}{High-level Security Policy Language}
\newacronym{mspl}{MSPL}{Medium-level Security Policy Language}
\newacronym{nsf}{NSF}{Network Security Function}
\newacronym{cti}{CTI}{Cyber Threat Intelligence}
\newacronym{ioc}{IoC}{Indicator of Compromise}
\newacronym{nfv}{NFV}{Network Functions Virtualization}
\newacronym{sdn}{SDN}{Software-Defined Networking}
\newacronym{stix}{STIX}{Structured Threat Information Expression}
\newacronym{llm}{LLM}{Large Language Model}
\newacronym{nlp}{NLP}{Natural Language Processing}
\newacronym{smt}{SMT}{Satisfiability Modulo Theories}
\newacronym{waf}{WAF}{Web Application Firewall}
\newacronym{crs}{CRS}{Core Rule Set}
\newacronym{tosca}{TOSCA}{Topology and Orchestration Specification for Cloud Applications}
\newacronym{clips}{CLIPS}{C Language Integrated Production System}
\newacronym{scm}{SCM}{Security Capability Model}
\newacronym{xml}{XML}{eXtensible Markup Language} 

\newcommand{\refeng}{{RefinementEngine}\xspace}
\newcommand{\cti}{\gls{cti}\xspace}
\newcommand*\circled[1]{%
  \smash{\tikz[baseline=(char.base)]{
    \node[
      circle,
      draw,
      line width=0.4pt,
      inner sep=0pt,
      outer sep=0pt,
      minimum size=1.0em,
      font=\scriptsize
    ] (char) {#1};
  }}%
}

\usepackage{todonotes}

%% file: sections/introduction.tex
In \glspl{soc}, a typical task for analysts and security engineers is to map security intents into concrete enforcement actions to protect organisational assets and reduce the impact of cyberattacks~\cite{socstudychallenges}. A common approach is to express security requirements at a high level using intent-oriented languages, which offer a convenient abstraction for governance and reasoning~\cite{securitypolicyspecification,surveypolicylanguages}. However, high-level intents are not directly deployable. Operational networks comprise heterogeneous appliances, each supporting distinct security controls and configuration primitives. Even when the same logical intent is required, it may need to be enforced through different mechanisms depending on device capabilities, and it may only be enforceable at specific points that observe the relevant traffic~\cite{securitypolicyspecification,translationsecurityfunctions}. 
Policy allocation must therefore account for device-specific control support, which determines whether a device can enforce a given policy and in what form, as well as reachability constraints imposed by the network topology and traffic paths.

When such issues are addressed manually, the refinement and allocation process is rarely straightforward in production environments; it is error-prone, time-consuming, and a recurring source of misconfigurations~\cite{firewallconfigurationerrors}, as well as inconsistencies and overheads in large networks~\cite{policyanomaliesfirewall,anomaliesfirewallrules}.
This challenge is further amplified by the pace at which enforced intents must be updated. Networks and deployments evolve (e.g., new segments and devices, changing routes); security requirements change in response to emerging threats and organizational needs. In many organizations, management workflows are fragmented: intents are specified separately from enforcement, device-level configurations are written or adapted by hand, and validation often occurs at later stages through troubleshooting~\cite{socstudychallenges,rfcfirewallpolicy}. 

The goal of this research is to address a crucial bottleneck: defining methods to refine intents into deployable, control-specific configurations that are consistent with the actual network and its constraints~\cite{securitypolicyspecification,translationsecurityfunctions}, and maintain intents correctly enforced when operational conditions evolve. 
Existing approaches typically address only parts of this problem, such as policy specification, translation, anomaly analysis, or deployment validation, but do not jointly automate the entire pipeline while accounting for both topology-dependent reachability and heterogeneous device-control support.

To this purpose, this paper introduces \refeng, a system that automates the refinement and allocation of network filtering policies over a custom network. Its main contribution is a capability-aware and topology-aware refinement process that enforces and maintains intents with a minimum number of devices, while preserving reachability and accounting for control-specific constraints. 
From these decisions, \refeng deterministically generates the low-level configurations to deploy, through a model-based process.

The major achievement of our research is in the level of autonomy of the refinement process. 
When \refeng receives data that has not been previously encountered (e.g., new malware types characterized by different attributes or attack vectors that were not previously known), it adapts its enforcement of intent in the new context.
That is, the refinement process is not statically defined by a fixed set of data types known in advance. \refeng dynamically adapts the models so that, whenever a new type emerges, it correctly operates in the new context.

In this work, we consider a realistic upstream source of enforcement variations: \cti reports. 
In the proposed end-to-end instantiation, an AI-based module extracts security-relevant, configurable entities (i.e., \glspl{ioc}) from \cti and provides them as input artifacts for policy refinement and allocation. This approach makes the pipeline incrementally adaptable to previously unseen artifacts: when a new \cti contains network-relevant entities not observed in earlier reports, these are incorporated into the knowledge extracted by this module and processed by subsequent stages without disrupting execution. This enables an adaptive pipeline that transforms threat-relevant information into deployable network enforcement.

The proposed approach has been validated through real-world use cases, where \refeng managed packet and web filtering policies and updated them to address threats described in CTI reports, demonstrating both correctness and practical applicability. 
The evaluation indicates that networks managed with \refeng should significantly reduce expert effort and the risk of misconfiguration when implementing and maintaining network authorization policies throughout their lifecycles.

The remainder of the paper is organized as follows: Section~\ref{sec:background} introduces the base background knowledge for this contribution; Section~\ref{sec:implementation} describes the \refeng architecture and operational workflow; Section~\ref{sec:validation} presents the use cases and workflow evaluation; Section~\ref{sec:related} presents related works from the literature; finally, Section~\ref{sec:conclusions} concludes and outlines future directions.

%% file: sections/background.tex
This section introduces the foundational concepts and technologies required to understand the proposed work.

\subsection{Domain-specific issues}
This research addresses two core problems concerning the enforcement of security policies in heterogeneous networks: policy allocation and policy refinement.

\textbf{Policy allocation} is the process of determining where a high-level security requirement should be enforced within a network, i.e., selecting the enforcement points that must be configured so the policy applies to the intended traffic. In practice, allocation must account for reachability and visibility constraints imposed by topology and routing (e.g., the enforcing device must be on the relevant traffic paths), as well as feasibility constraints imposed by the supported security controls (e.g., the device must provide the enforcement primitives required by the policy). An effective allocation achieves complete policy coverage while avoiding unnecessary configuration, for example, by minimising the number of configured devices or consolidating enforcement across shared path segments.

\textbf{Policy refinement} is the process of translating a policy from an abstract, intent-oriented form into progressively more concrete representations that can be deployed on security controls. In this context, high-level policies are stated through \glspl{hspl} and express security intent independently of specific enforcement technologies; conversely, medium-level policies are expressed through \glspl{mspl} and capture the security artifacts and parameters required to realize that intent on compatible controls. Refinement resolves high-level specifications (e.g., abstract subjects/objects, services, and actions) into medium-level conditions and actions, and maps policy semantics onto the configuration primitives supported by the target enforcement mechanisms. In heterogeneous environments, refinement is constrained by device capabilities and target languages; therefore, it must include proper checks of enforceability and select compatible control features. As a result, it produces intermediate, control-aligned policy artifacts that can then be deterministically translated into low-level configurations.

\subsection{Solutions for implementation}
\label{background:implementation}
To implement \refeng, the following technologies were employed at different stages of the execution pipeline, and their use will be discussed in detail in Section~\ref{sec:implementation}.

\textbf{\gls{clips}}\footnote{\url{https://www.clipsrules.net/}} is a rule-based expert system environment designed to represent domain knowledge as facts and production rules, and to derive conclusions through forward-chaining inference. In this work, \gls{clips} provides a declarative substrate for encoding network and policy artifacts (e.g., topology-derived facts, device properties, and capability constraints) and for performing refinement decisions via rule firing rather than ad hoc procedural logic. This choice supports modularity, allowing new reasoning steps to be added as rules rather than modifying the workflow; in parallel, the decision process remains explicit and inspectable (i.e., derived artifacts can be traced back to the facts and rules that produced them). Such properties are deemed beneficial when refinement must be auditable and reproducible. \gls{clips} also supports this work with its stringent syntax: given the criticality of the operative domain, oversights are unacceptable; \gls{clips} integrates a highly reliable compilation system that raises detailed errors when the underlying system attempts actions beyond what is specifically allowed.

\textbf{\gls{tosca}}\footnote{\url{https://docs.oasis-open.org/tosca/}} is a standard language for describing systems as typed topologies of components and relationships, complemented by declarative management and lifecycle information. In this work, TOSCA is used as a structured notation to represent the input network (e.g., nodes, connectivity, and device properties) in a machine-processable and semantically explicit form. Compared to ad hoc JSON, a TOSCA-based representation provides a reusable type system and well-defined relationship semantics, enabling validation and consistent interpretation across tools and implementations; furthermore, it supports systematic reasoning about reachability and device placement.

\textbf{Security capabilities} capture, in a vendor-neutral form, the enforcement features exposed by technical security controls (e.g., firewalls) and the configuration primitives through which they can be expressed. The \textbf{\gls{scm}}, as referenced from the relevant literature~\cite{basile2024formalmodelsecuritycontrols}, formalizes this notion by abstracting what a security control can enforce through capabilities for rule conditions, actions, and (when applicable) events; these capabilities are complemented by policy-level semantics such as default actions and resolution strategies. By separating an information model of generic policy concepts from a data model that enumerates concrete capability types for families of controls (e.g., filtering and channel protection), \gls{scm} enables systematic reasoning about how to compare and enforce policies across different devices. Moreover, by associating capabilities with translation details specific to security controls, the model supports a model-driven translation from abstract policies to device-specific configuration settings; hence, reliance on ad hoc, hard-coded refiners and translators is reduced in automated enforcement workflows.

\textbf{Agents.} Contemporary AI agents are commonly characterized by three main features~\cite{AI_agent}: they are designed to pursue specific tasks, they operate on text and multimodal data as inputs and outputs, and they rely on \glspl{llm} for reasoning. The use of \glspl{llm} has been identified as a controversial issue in cybersecurity, as evidenced by the literature \cite{Ares_LLM}. The inherent challenges of \gls{llm} functionality, including determinism and task scope, have been identified as significant concerns. In this study, they are addressed by integrating best practices from the \gls{clips} language and intrinsic programming syntax, leveraging the semantics of \cti, and enhancing the proposed solution with the pertinent agent-like features of a neurosymbolic approach, which are advantageous in the domain. Moreover, following state-of-the-art best practices, deterministic \gls{llm} inference is approximated by mitigating both hardware and software nondeterminism. Greedy decoding and explicit device placement further reduce output variability \cite{song-etal-2025-good}. Despite its inherent heuristic nature, this configuration is effective in ensuring reproducible results.

\subsection{Security controls for validation}
To validate the \refeng, as described in Section~\ref{sec:validation}, we consider two widely used security controls: iptables and ModSecurity. Their complementary nature demonstrates the effectiveness of the proposed work in adapting to different filter types across varying levels of granularity.

\textbf{iptables\footnote{\url{https://www.netfilter.org/projects/iptables/}}} is an open-source firewall utility for Linux that configures the kernel’s Netfilter packet-processing framework. It is commonly used to implement IP-based filtering policies by matching packet-level and flow-level attributes (e.g., source/destination IP addresses, transport protocols, ports, interfaces, and connection state) and applying actions such as accept, drop, reject, or log. In this work, iptables serves as the target security control for network-layer filtering, with enforcement expressed in terms of packet and flow properties.

\textbf{ModSecurity\footnote{\url{https://github.com/owasp-modsecurity/ModSecurity}}} is an open-source \gls{waf} engine for HTTP(S) traffic inspection, typically deployed through web-server connectors (e.g., Apache and Nginx). It applies a rule-based language to analyze requests and responses and trigger enforcement actions, is therefore a natural enforcement choice for web-filtering policies that depend on application-layer visibility (e.g., URLs, headers, parameters, and payload patterns), and it can be coupled with shared rule sets such as the OWASP \gls{crs}\footnote{\url{https://owasp.org/www-project-modsecurity-core-rule-set/}} to provide a baseline of generic web-attack detection that can be tailored to the protected application.

%% file: sections/implementation.tex
The main contribution of this research is \refeng, a solution for automatic policy allocation and refinement that bridges the gap between intent-oriented policy specification and deployable enforcement in different networks. It achieves this by converting high-level requirements into device-specific configurations while respecting topology reachability and device constraints.

\begin{figure}[tb]
    \centering
    \includegraphics[width=1.00\linewidth]{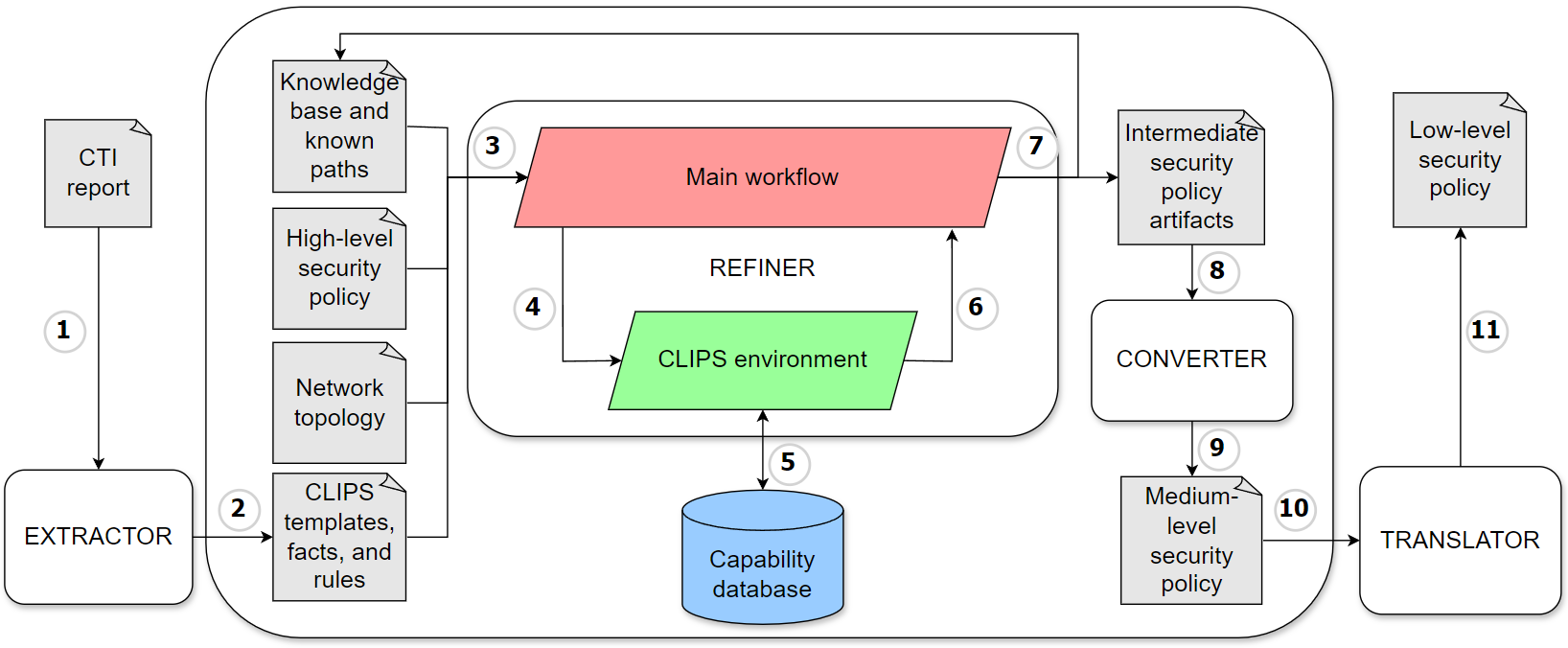}
    \caption{Architecture and end-to-end workflow of RefinementEngine}
    \label{fig:refeng}
\end{figure}

The architecture of \refeng, presented in Figure~\ref{fig:refeng}, is organized as a pipeline of four components. First, the \textit{Extractor} processes natural-language \cti reports and produces network and security entities that the refinement logic can consume. Second, the \textit{Refiner} performs the core reasoning step: it combines the extracted entities, the input \gls{hspl} intent, and the network model to determine feasible enforcement decisions and derive intermediate policy artifacts. Third, the \textit{Converter} processes the intermediate artifacts and outputs \gls{mspl} policies, specifically one policy file per enforcement device. Finally, the \textit{Translator} converts each \gls{mspl} policy into low-level rules in the native configuration language of the designated security control.

A key design principle of the \refeng is to maximize deterministic approaches to drive enforceability and translation.
In particular, device feasibility is evaluated using a well-established security capability model. Also, translation from \gls{mspl} to low-level language is performed via a model-based translation procedure grounded in the \gls{scm}, in both cases without the use of generative AI. 
This design yields reproducible outputs and avoids introducing nondeterminism or best-effort heuristics whenever AI-based solutions can be avoided.

\subsection{Workflow overview}
\label{implementation:workflow}
The workflow in Figure~\ref{fig:refeng} starts from a natural-language \cti report provided to the Extractor \circled{1}. The Extractor identifies security and network-relevant entities \circled{2} (e.g., \glspl{ioc}) and incrementally structures them as \gls{clips} knowledge (templates, facts, and rules). 
The Refiner receives this knowledge together with the input \gls{hspl} policy and the \gls{tosca} network topology \circled{3}. During initial execution, these inputs are supplied externally and used to derive the relevant network paths and device inventory; in subsequent executions, this inventory may be reused as precomputed inputs (i.e., a knowledge base) to reduce execution overhead, unless changes in topology or available devices require recomputation. Within \gls{clips}, the Refiner derives the relevant network and device attributes \circled{4}, retrieves the security capabilities needed to realize the \cti intent \circled{5}, and applies the \gls{hspl} intent to compute feasible enforcement decisions \circled{6}. The Refiner outputs intermediate artifacts (selected enforcement devices, security capabilities, and parameters) \circled{7}, which are then passed to the Converter \circled{8}. The Converter constructs the corresponding \gls{mspl} policy \circled{9} and forwards it to the Translator \circled{10}. The Translator generates low-level filtering rules by mapping security capabilities to the target controls’ configuration primitives \circled{11}. The resulting rules are ready for deployment on the candidate devices, as will be validated in Section~\ref{sec:validation}.

\subsection{Extractor}
\label{implementation:extractor}

The \textbf{Extractor} serves as the initial component of the proposed framework, responsible for parsing \glspl{cti} documentation and deriving network and security artifacts. \gls{cti} reports typically describe malware behavior and \glspl{ioc} associated with adversarial network activity. Leveraging \gls{nlp} and \glspl{llm} inference, the Extractor semantically processes this unstructured text to identify and isolate security-relevant entities into \gls{clips} knowledge.

The strong influence of \gls{clips} language is particularly crucial in this phase. Recent literature has highlighted that \glspl{llm} can be unreliable when processing security-critical data~\cite{llm_unreliable}. To counter this issue, as presented in Section~\ref{sec:background}, \gls{clips}' syntax manages stringent constraints on what the \gls{llm} is allowed to do, which are placed upon an already existent and tailored substrate of guardrails; furthermore, this programming language is highly involved in the symbolic branch of Artificial Intelligence (i.e., an approach that uses explicit, human-readable symbols, rules, and logic to represent knowledge and solve problems). It has been demonstrated how it achieves high accuracy, interpretability, and robust reasoning when combined with other approaches for Cybersecurity~\cite{NeSyAIcyber,NeSyAIdefend}. Furthermore, the underlying heuristics to approximate determinism for \glspl{llm} also strengthen the countermeasures that this work includes to mitigate \glspl{llm}' downsides as much as the current state of the art offers. 
The complete structure of the Extractor and its approximation of deterministic inference is thoroughly explained and validated in parallel work from the literature~\cite{nostro_agente}.

Listing~\ref{lst:ctiexample} presents a typical natural language input that the Extractor is capable of processing. It consists of a snippet from a \gls{cti} report that lists a malicious IP address as an \gls{ioc} observed for a malware instance. Listing~\ref{lst:clipsexample} reports the consequent output: the \gls{ioc} is structured into \gls{clips} templates and facts that will be leveraged by the refinement logic to assess the entities involved in the security intents.

\begin{lstlisting}[label={lst:ctiexample},caption={Original natural language CTI - Example},captionpos=b,frame=tb,float=tb]
...The malware compromises vulnerable hosts and sends requests from the address 1.2.3.4...
\end{lstlisting}

\begin{lstlisting}[label={lst:clipsexample},caption={Extracted CLIPS knowledge - Example},captionpos=b,frame=tb,float=tb]
"templates": [
  "(deftemplate entity (slot source-ip-address (type STRING)) ... )"
],
"facts": [
  "(entity (source-ip-address \"1.2.3.4\") ... )"
]
\end{lstlisting}

\subsection{Refiner}
\label{implementation:refiner}
The \textbf{Refiner} is the central component of this contribution and maps security intents onto the input network topology. It receives an input \gls{hspl} policy and derives the intermediate network artifacts to apply the provided security intents over a network topology.

\begin{lstlisting}[label={lst:hsplexample},caption={HSPL policy - Example},captionpos=b,frame=tb,float=b]
<hspl id=<identifier of the original HSPL policy>>
  <subject>Endpoint_A</subject>
  <action>is not authorized to access</action>
  <object>Endpoint_B</object>
</hspl>
\end{lstlisting}

Listing~\ref{lst:hsplexample} shows an example \gls{hspl} policy that \refeng can process, in combination with the \gls{clips} knowledge (Listing~\ref{lst:clipsexample}). The policy is identified by a unique \texttt{id} (here left as a placeholder) and specifies a \texttt{subject}, \texttt{action}, and \texttt{object}. 
In the proposed workflow, these entities are resolved against the network topology and the \gls{clips} knowledge to instantiate enforceable filtering constraints. Objects and subjects can be further detailed with attributes. 

To avoid redundant computation across successive runs, the Refiner first compares the incoming \gls{hspl} and topology-related inputs against a pre-computed knowledge base of intents, paths, and device descriptions. When an overlap is detected, either in the requested intents or in the relevant network details, the Refiner reuses prior results and skips the associated processing stages, thereby improving overall efficiency. As a clarifying example, if the input network remains unchanged from previous computations, all device details are retrieved from the knowledge base rather than gathered by exploring the \gls{tosca} network topology.

If no reusable knowledge is available, the Refiner performs a full computation round. It injects into a \gls{clips} environment both the \cti-derived knowledge produced by the Extractor and the intents specified in the \gls{hspl}. 
It then analyzes the network topology to derive all the possible paths, i.e., all feasible routes between the communicating parties, so that enforcement decisions cover any route permitted by the modeled connectivity; it also collects the security controls supported by the devices along those paths. 

Based on the policy intents, the Refiner asserts the \gls{clips} facts related to the intents and queries the database of security capabilities to retrieve the ones \textit{needed} to enforce them. By analysing the security-control support provided by devices along paths, the Refiner determines the \textit{available} capabilities and selects one or more minimal enforcement sets, i.e., the smallest number of devices that can enforce the intent across all applicable paths.
The assumption is that, to block traffic, it is sufficient to identify a single device that can deny forwarding on all paths to the destinations, thereby accounting for the enforced intent. 
Further optimization is possible, as demonstrated in the literature~\cite{valenza2017formal}; however, this work focuses on enforcing intents with a minimum number of controls, without pursuing more general or sophisticated optimizations, which will be integrated as future work.

During this step, the Refiner also accounts for shared sub-paths, so that selecting an enforcement point common to multiple paths can provide coverage without redundant configuration. For instance, if two paths share a common device with suitable security controls, choosing that device ensures coverage for both paths. Thus, every modeled path traverses at least one enforcement device capable of implementing the required control, preventing policy bypass. If no such set exists (e.g., at least one feasible path contains no device supporting the required control), the Refiner flags the intent as unenforceable under the current topology and capability constraints.

Upon completion, the Refiner outputs the selected enforcement devices, the candidate control for each device, and the security capabilities required to instantiate the filtering rules at later stages. Finally, the Refiner updates the knowledge base with the newly derived intents, paths, and device information, enabling subsequent executions to reuse results whenever the inputs overlap.

Listing~\ref{artifactsexample} presents the Refiner's partial output when provided with the \gls{hspl} policy and \gls{clips} knowledge in Listings~\ref{lst:clipsexample} and~\ref{lst:hsplexample}, as well as a custom \gls{tosca} network. It is a list of rule artifacts, one for each filtering rule enforced on a device. The \texttt{hsplid}, \texttt{device}, and \texttt{nsf} keys (which have placeholder values) are respectively: the \gls{hspl} policy to which the artifacts refer; the unique identifier of the network device to be configured for enforcing the intent; the candidate configuration language for a security control supported by the device. The \texttt{capabilities} list contains all security capabilities and details that the low-level rule must reflect when the Translator translates it into the control's configuration language.

\begin{lstlisting}[label={artifactsexample},caption={Intermediate network artifacts - Example},captionpos=b,frame=tb,float=tb]
[
  {
    "hsplid": <identifier of the original HSPL policy>,
    "device": <identifier of the network device>,
    "nsf": <name of the candidate security control>,
    "capabilities": [
      ...
      {
        "capability": "IpSourceAddressConditionCapability",
        "detail": "1.2.3.4"
      },
      ...
    ]
  },
  ...
]
\end{lstlisting}

\subsection{Converter}
\label{implementation:converter}
The \textbf{Converter} builds the \gls{mspl} policies from the intermediate artifacts produced by the Refiner. These artifacts specify, for each selected device, the chosen security control to configure and the set of capability-specific parameters required to instantiate the filtering rules. During generation, the Converter aggregates all rules by target device and produces a single \gls{mspl} policy file per device, which contains a self-contained policy document that includes all rules it must enforce. In producing this document, the Converter not only serializes the artifacts into \gls{xml}, but also normalizes parameter values into the structured formats expected by the \gls{mspl} schema (e.g., representing address ranges and sets of states in a canonical form). The resulting \gls{mspl} files are then provided to the Translator for conversion into low-level configurations.

Starting from Listing~\ref{artifactsexample}, Listing~\ref{msplexample} shows how the Converter generates the corresponding \gls{mspl} policy. The \texttt{nsfName} and \texttt{id} fields carry the designated security control and unique policy identifier, and are again reported with placeholder values; the \gls{xml} object \texttt{ipSourceAddressConditionCapability} is the notation for the security capability about source IP addresses, processable by the Translator (the \texttt{exactMatch} operator means that the capability refers to a single address, as the Translator supports unions and ranges of addresses as well).

\begin{lstlisting}[label={msplexample},caption={MSPL policy - Example},captionpos=b,frame=tb, float=tb]
<?xml version='1.0' encoding='utf-8'?>
<policy ... nsfName=<name of the candidate security control>>
  <rule id=<identifier of the original HSPL policy>>
    ...
    <ipSourceAddressConditionCapability operator="exactMatch">
      <capabilityIpValue>
        <exactMatch>1.2.3.4</exactMatch>
      </capabilityIpValue>
    </ipSourceAddressConditionCapability>
    ...
  </rule>
</policy>
\end{lstlisting}

\subsection{Translator}
\label{implementation:translator}
The \textbf{Translator} represents the final stage of the enforcement pipeline, converting \gls{mspl} policies into deployable configurations. Concretely, it receives an \gls{mspl} policy expressed in \gls{xml}, generated by the Converter in the previous stage and annotated with the target security control, and produces low-level rules in the native language of that control (e.g., \texttt{iptables} commands). 

A key design choice of the Translator is that translation is performed through a model-based procedure grounded in the \gls{scm}, as presented in Section~\ref{background:implementation}. For each security control, the Translator relies on a specific capability schema and a deterministic mapping from capability instances (including operators and typed values) to the corresponding configuration primitives of the low-level configuration language. As a consequence, translation is fully deterministic: given the same \gls{mspl} policy input, the Translator always produces the same low-level output for each control. 

Listing~\ref{lowlevelexample} is a representative output of the Translator when provided with the \gls{mspl} policy from Listing~\ref{msplexample}. For this example, it is assumed that the target security control was \texttt{iptables}, showing how \gls{mspl} capability instances are rendered as command-line arguments. Note that the low-level rule contains only the information extracted from the CTI, i.e., the iptables notation for source IP addresses, inherited from the initial \cti as intended by the workflow.
\begin{lstlisting}[label={lowlevelexample},caption={Low-level policy - Example},captionpos=b,frame=tb,float=tb]
iptables ... -s 1.2.3.4 ...
\end{lstlisting}

%% file: sections/validation.tex
The outcomes of this research were evaluated to assess the correctness and applicability of \refeng. The goal was to verify that, given a real \cti report and a network with multiple paths and devices between two endpoints, the pipeline schematized in Figure~\ref{fig:refeng} derives the appropriate filtering policies to enforce the intents implied by the \cti. Moreover, such policies are deemed satisfactory if they are also correct with respect to the candidate security control's configuration language; finally, according to our simplified approach hypothesis, they must be enforced on the minimum number of devices to secure all relevant paths.
The evaluation also illustrates how the proposed pipeline reduces reliance on manual analysis and intervention by automatically deriving rules from high-level intent and adapting to new knowledge, thereby reducing opportunities for human error and omissions.
Tests were conducted on real scenarios described in the quarterly malware reports from the Center of Internet Security\footnote{\url{https://portal.cisecurity.org/}}, taking reports from the last three years (2023-2025) and providing them sequentially to the \refeng with different network topologies. In the following sections, two sample scenarios from those reports are presented in detail to simulate the flow of incoming reports. One of these reports concerns IP packet filtering, and the other concerns web filtering, as they are the simplest and most straightforward cases.

\subsection{Test scenario 1 - IP packet filtering}
\label{validation:iptest}
The first test scenario involves \emph{CoinMiner}, a malware listed in the CIS Top 10 Malware report for the third quarter of 2024\footnote{\url{https://portal.cisecurity.org/insights/articles/top\%2010\%20malware\%20q3\%202024}}. The report briefly describes the nature of the malware and identifies IP address \texttt{80.71.158.96} as an \glspl{ioc}. 

\begin{figure}[tb]
    \centering
    \includegraphics[width=0.80\linewidth]{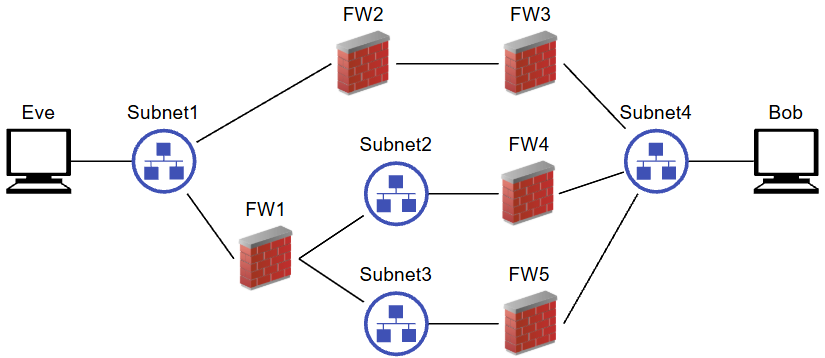}
    \caption{Topology for Test scenario 1}
    \label{fig:testscenario1}
\end{figure}

Referring to the workflow in Figure~\ref{fig:refeng}, the Extractor processes the natural-language report on \emph{CoinMiner}, extracts relevant network artifacts and \glspl{ioc}, and passes them to the Refiner as \gls{clips} structures, presented in Listing~\ref{lst:clipsTest1}.

\begin{lstlisting}[label={lst:clipsTest1},caption={CLIPS knowledge for Test Scenario 1},captionpos=b,frame=tb]
"templates": [
  "(deftemplate entity (slot destination-ip-address (type STRING)))"
],
"facts": [
  "(entity (destination-ip-address \"80.71.158.96\"))"
]
\end{lstlisting}

The operative network topology is represented in Figure~\ref{fig:testscenario1}. In this scenario, the previous knowledge base and paths are empty, and the high-level filtering policy is outlined in Listing~\ref{lst:hsplTest1}. 

\begin{lstlisting}[label={lst:hsplTest1},caption={HSPL for Test Scenario 1},captionpos=b,frame=tb]
<hspl id="hspl1">
  <subject>Eve</subject>
  <action>is not authorized to access</action>
  <object>Bob</object>
</hspl>
\end{lstlisting}

In this setup, Eve is the compromised host and is assigned the malicious IP address specified in the report; Bob is a normal user with a generic IP address (i.e., \texttt{172.19.0.3}), who could potentially be Eve's target.

The Refiner receives all inputs described above and, leveraging the \gls{clips} environment, identifies all paths between Eve and Bob, as well as the minimal set of devices and security capabilities to be configured to achieve complete protection coverage. The paths are presented in Listing~\ref{lst:pathsTest1}.

\begin{lstlisting}[label={lst:pathsTest1},caption={Paths identified for Test Scenario 1},captionpos=b,frame=tb]
"hspl1": [
    ["FW1", "Subnet2", "FW4"],
    ["FW1", "Subnet3", "FW5"],
    ["FW2", "FW3"]
]
\end{lstlisting}

Figure~\ref{lst:msplTest1} presents the list of devices to be configured, the security controls (referenced to as \texttt{nsf} in the JSON structure), and security capabilities necessary for each policy.

\begin{figure}
\centering
\begin{minipage}{0.49\linewidth}
\vspace{0pt}
\begin{lstlisting}[frame=tb]
[
  {
    "hsplid": "hspl1",
    "device": "FW1",
    "nsf": "IpTables",
    "capabilities": [...]
  },
  {
    "hsplid": "hspl1",
    "device": "FW1",
    "nsf": "IpTables",
    "capabilities": [...]
  },
\end{lstlisting}
\end{minipage}
\hfill
\begin{minipage}{0.49\linewidth}
\vspace{0pt}
\begin{lstlisting}[frame=tb]
  {
    "hsplid": "hspl1",
    "device": "FW3",
    "nsf": "IpTables",
    "capabilities": [...]
  },
  {
    "hsplid": "hspl1",
    "device": "FW3",
    "nsf": "IpTables",
    "capabilities": [...]
  }
]
\end{lstlisting}
\end{minipage}
\caption{Artifacts identified by RefinementEngine for Test Scenario 1}
\label{lst:msplTest1}
\end{figure}

It is noteworthy that \texttt{FW1} is identified as a suitable device for configuration because it is shared between the first two paths. The third path is distinct and comprises two configurable devices; however, \texttt{FW3} is selected over \texttt{FW2} because the latter does not support any security controls suitable for this IP-based filtering scenario. Furthermore, for each device, the Refiner generates two groups of artifacts; since communication between Eve and Bob is bi-directional by construction, \texttt{iptables} requires one filtering rule per direction.

The list of security capabilities includes all features needed to enforce low-level security policies when translating from \gls{mspl} to iptables for \texttt{FW1} and \texttt{FW3}. For example, it contains a security capability representing the source IP address of the transmitted data and another that specifies the destination IP address.

These artifacts are provided to the Converter, which constructs the \gls{mspl} policy for each device and forwards it to the Translator, which in turn translates the \gls{mspl} policies into iptables-specific rules, as shown in Listing~\ref{lst:lowLevelTest1}.

\begin{lstlisting}[label={lst:lowLevelTest1},caption={Filtering rules generated by RefinementEngine for FW1 and FW3},captionpos=b,frame=tb]
iptables -A FORWARD -m conntrack --ctstate NEW,ESTABLISHED -s 80.71.158.96 -d 172.19.0.3 -j DROP
iptables -A FORWARD -m conntrack --ctstate ESTABLISHED,RELATED -s 172.19.0.3 -d 80.71.158.96 -j DROP
\end{lstlisting}

Performing ICMP requests from Eve to Bob and sniffing the traffic reveals an unfiltered flow of requests and replies (see Listing~\ref{lst:trafficBeforeTest1}). The traffic was observed both on \texttt{FW1}'s interface towards Eve's subnet and \texttt{FW3}'s interface towards \texttt{FW2}.

\begin{lstlisting}[label={lst:trafficBeforeTest1},caption={IP traffic before filtering rules - Test Scenario 1},captionpos=b,frame=tb]
... eth1  In  IP 80.71.158.96 > 172.19.0.3: ICMP echo request ...
... eth1  Out IP 172.19.0.3 > 80.71.158.96: ICMP echo reply ...
... eth1  In  IP 80.71.158.96 > 172.19.0.3: ICMP echo request ...
... eth1  Out IP 172.19.0.3 > 80.71.158.96: ICMP echo reply ...
... eth1  In  IP 80.71.158.96 > 172.19.0.3: ICMP echo request ...
... eth1  Out IP 172.19.0.3 > 80.71.158.96: ICMP echo reply ...
\end{lstlisting}

After enforcing the generated iptables rules on \texttt{FW1} and \texttt{FW3}, traffic was again analyzed at the same points. It can be noted that Eve's requests are never answered, as they are dropped by \texttt{FW1} and \texttt{FW3} (see Listing~\ref{lst:trafficAfterTest1}). The logs on Eve's terminal also specify that all sent packets are lost.

\begin{lstlisting}[label={lst:trafficAfterTest1},caption={IP traffic after filtering rules - Test Scenario 1},captionpos=b,frame=tb]
... eth1  In  IP 80.71.158.96 > 172.19.0.3: ICMP echo request ...
... eth1  In  IP 80.71.158.96 > 172.19.0.3: ICMP echo request ...
... eth1  In  IP 80.71.158.96 > 172.19.0.3: ICMP echo request ...
\end{lstlisting}

These results demonstrate that, starting from a \cti about a malware that operates through compromised IP addresses, \refeng is capable of (i) extracting and recognizing the IP-layer nature of the relevant \glspl{ioc}, (ii) identifying the useful security controls available in the network, depending on the location of the endpoints, and (iii) generating the necessary low-level filtering rules for such controls, ready to be deployed in a minimum set of devices. Hence, malicious traffic exhibiting the characteristics described in the original \cti is correctly filtered by the generated downstream rules without any manual intervention.

\subsection{Test scenario 2 - Web filtering}
\label{validation:webtest}

The second test scenario, provided some iterations after the previous one, involves a domain-based filtering use case, in which a client attempts to access a web server through a specific domain name. Malware \emph{NanoCore} is referenced from the CIS Top 10 Malware report for the third quarter of 2024\footnote{\url{https://portal.cisecurity.org/insights/articles/top\%2010\%20malware\%20q3\%202024}}. Domain \path{hadleyshope.3utilities.com} is reported as malicious among the \glspl{ioc} observed for \emph{NanoCore}.

\begin{figure}
    \centering
    \includegraphics[width=0.85\linewidth]{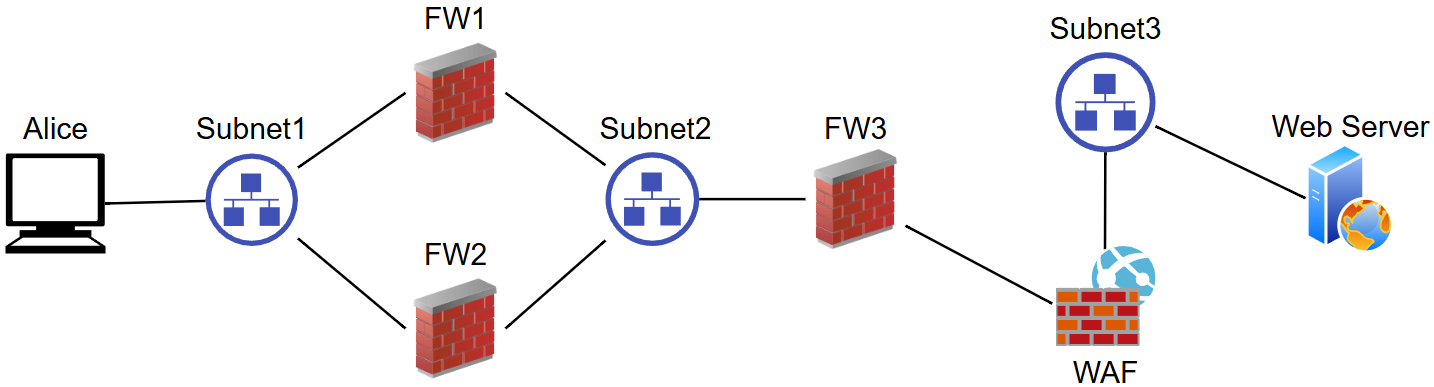}
    \caption{Topology for Test scenario 2}
    \label{fig:testscenario2}
\end{figure}

In this situation, \refeng is not initially equipped with the necessary knowledge, and the new \gls{ioc} about web domains would cause the system to crash (see Listing~\ref{lst:clipsTest1}). Still, given the adaptive nature of \gls{clips}, \refeng recognizes a new type of artifact in the \cti report, which does not map to existing knowledge (i.e., \texttt{destination-ip-address}). Consequently, \gls{clips} templates are updated to enable the environment to rigorously process the new indicators. 
Referring again to the workflow in Figure~\ref{fig:refeng}, the Extractor processes the natural-language report on \emph{NanoCore} and extracts the domain-based \glspl{ioc} and relevant network artifacts, which are then passed to the Refiner as CLIPS structures, reported in Listing~\ref{lst:clipsTest2}. 

\begin{lstlisting}[label={lst:clipsTest2},caption={CLIPS knowledge for Test Scenario 2},captionpos=b,frame=tb]
"templates": [
  "(deftemplate entity 
    (slot destination-ip-address (type STRING))
    (slot url (type STRING)))"],
"facts": [
  "(entity (url \"hadleyshope.3utilities.com\"))"]
\end{lstlisting}

It is noteworthy that the new knowledge structure continues to recognize future \glspl{ioc} involving remote IP addresses and, with the newly incoming report, seamlessly learns about URL components.
The operative network topology is represented in Figure~\ref{fig:testscenario2}. Also in this scenario, the previous knowledge base and paths are empty, and the high-level filtering policy is outlined in Listing~\ref{lst:hsplTest2}.

\begin{lstlisting}[label={lst:hsplTest2},caption={HSPL for Test Scenario 2},captionpos=b,frame=tb]
<hspl id="hspl2">
  <subject>Alice</subject>
  <action>is not authorized to access</action>
  <object>WebServer</object>
</hspl>
\end{lstlisting}

In this setup, Alice is the client host attempting to reach the Web Server. The Web Server exposes its service on two domains: a benign domain, \path{allowed.utilities.com}, and a compromised domain, \path{hadleyshope.3utilities.com}, as reported in the \cti.

The topology includes three firewalls (\texttt{FW1}, \texttt{FW2}, and \texttt{FW3}) that operate at the network layer. The \gls{waf}, positioned on the only ingress path to \texttt{Subnet3} and the Web Server, is the sole enforcement point that can inspect incoming HTTP requests and apply domain-based filtering.

The Refiner receives these inputs and identifies all paths between Alice and the Web Server, as presented in Listing~\ref{lst:pathsTest2}.

\begin{lstlisting}[label={lst:pathsTest2},caption={Paths identified by RefinementEngine for Test Scenario 2},captionpos=b,frame=tb]
"hspl2": [
  ["Subnet1", "FW1", "Subnet2", "FW3", "WAF", "Subnet3"],
  ["Subnet1", "FW2", "Subnet2", "FW3", "WAF", "Subnet3"]
]
\end{lstlisting}

Listing~\ref{lst:msplTest2} presents the list of devices, security controls, and security capabilities selected for the policy.

\begin{lstlisting}[label={lst:msplTest2},caption={Artifacts identified by RefinementEngine for Test Scenario 2},captionpos=b,frame=tb]
[
  {
    "hsplid": "hspl2",
    "device": "WAF",
    "nsf": "ModSecurity",
    "capabilities": [...]
  }
]
\end{lstlisting}

The \gls{waf} is chosen as the enforcement device because it protects all viable paths leading to the Web Server and supports application-layer inspection, which is necessary for the domain-based \glspl{ioc}. The list of security capabilities includes the features required to translate the derived \gls{mspl} policy into a ModSecurity rule for \texttt{WAF}, including the application-layer field representing the domain.

These artifacts are provided to the Converter that builds the \gls{mspl} policy for the chosen device and forwards it to the Translator, which in turn translates the \gls{mspl} policy into a ModSecurity-specific rule, as shown in Listing~\ref{lst:lowLevelTest2}.

\begin{lstlisting}[label={lst:lowLevelTest2},caption={Filtering rules generated by RefinementEngine for WAF},captionpos=b,frame=tb]
SecRule REQUEST_HEADERS:Host "@rx ^hadleyshope\.3utilities\.com$" \
  "deny, id:1"
\end{lstlisting}

Performing HTTP requests from Alice to the Web Server using both domains in the header reveals unfiltered behavior when no rule is enforced at the \gls{waf} level (see Listing~\ref{lst:trafficBeforeTest2}). Traffic sniffed on \texttt{FW3}'s interfaces confirms that both requests appear identical at the network layer (same source and destination IP addresses, same TCP destination port, and comparable payload sizes), indicating that \texttt{FW3} has no means to distinguish between the two domains.

\begin{lstlisting}[label={lst:trafficBeforeTest2},caption={HTTP responses before enforcing the ModSecurity rule},captionpos=b,frame=tb]
[allowed.utilities.com] HTTP response code: 200
[hadleyshope.3utilities.com] HTTP response code: 200
\end{lstlisting}

After enforcing the ModSecurity filtering rule on the \gls{waf}, the requests were repeated. As shown in Listing~\ref{lst:trafficAfterTest2}, requests carrying the \path{hadleyshope.3utilities.com} domain are rejected with a \texttt{403 Forbidden} response, while requests for \path{allowed.utilities.com} continue to be served normally.

\begin{lstlisting}[label={lst:trafficAfterTest2},caption={HTTP responses after enforcing the ModSecurity rule},captionpos=b,frame=tb]
[allowed.utilities.com] HTTP response code: 200
[hadleyshope.3utilities.com] HTTP response code: 403
\end{lstlisting}

Crucially, traffic sniffed on \texttt{FW3} during both phases is indistinguishable: the firewall observes TCP connections towards the Web Server with no visibility into the HTTP header embedded in the payload. The enforcement decision is therefore made entirely within the \gls{waf} at the application layer.

The results indicate that, for contexts involving malicious activity that can only be identified at the application layer (such as domain-based \glspl{ioc}), the \refeng is capable of (i) extracting and recognizing the application-layer characteristics of the \glspl{ioc}, (ii) selecting enforcement points that offer the necessary security controls, and (iii) generating the required low-level rules for these controls, all ready for deployment on the minimum number of devices. Therefore, connections toward domains listed as \glspl{ioc} in the original \cti are correctly blocked by the generated downstream rules.

To conclude, the correctness of the flow described in both test scenarios is ensured by automated mechanisms and does not require manual intervention; the surface for misconfiguration due to human intervention is thus highly limited. Also, new \glspl{ioc} might appear in other reports (e.g., HTTP payload, TCP port); the workflow, supported by \gls{clips}' inherent peculiarities, manages to integrate new types of indicators without disrupting the execution.

%% file: sections/related.tex
Several works from the literature address policy refinement in \gls{nfv} and \gls{sdn} environments.
Cheminod et al.~\cite{cheminod2019automaticrefinement} have presented a comprehensive methodology for refining and verifying access-control policies; it uses a dual model of policies and system configurations that can either synthesize a correct configuration or diagnose mismatches and suggest fixes.
In \gls{sdn}, a complementary line of work addresses enforcement robustness against dynamic updates and conflicting rules. Porras et al.'s FortNOX~\cite{porras2012securitynetworks} enforces authorization and security constraints at the controller to prevent conflicting flow-rule installations. At the same time, FlowGuard, by Hu et al.~\cite{hu2014flowguard}, detects and resolves firewall policy violations by analyzing flow-path spaces under evolving network state.
Relatedly, Schnepf et al. have proposed Synaptic~\cite{schnepf2017automated}, which automates the verification of security chains by translating chain specifications into formal models that can be checked via \gls{smt} solving or model checking.
These systems strengthen correctness and runtime safety properties in controller-driven networks. Still, they do not directly address our central objective of deriving minimal enforcement sets over heterogeneous devices under explicit reachability and control-support constraints.

In parallel, the current state of the art poses placement as a resource-driven and performance-driven optimization problem in \gls{nfv}/\gls{sdn} infrastructures.
Li et al.~\cite{li2016nsfsurvey} have surveyed network function placement strategies and their design trade-offs, while Demirci et al.~\cite{demirci2019nsfplacement} have reviewed virtual security functions and their placement in \gls{sdn}, highlighting open challenges and common optimization criteria.
Both surveys characterize placement primarily as selecting where to instantiate functions to satisfy resource and quality of service objectives under infrastructure constraints, typically assuming controller-orchestrated deployment of virtualized functions. As a result, placement is often decoupled from policy refinement: optimization decides where to deploy a function, and the policy configuration is assumed to be realizable once a function is placed.
In contrast, the \refeng focuses on configuring existing enforcement points for filtering policies; it jointly reasons over traffic reachability and device-specific control support to determine where policies can be enforced, and it produces device-aligned policy refinements ready for low-level deployment.

A relevant body of work employs abstraction via \gls{hspl} and \gls{mspl} to bridge user-oriented security intent with control-specific enforcement.
Montero et al.~\cite{montero2015virtualized} embed this abstraction within an NFV-enabled security-at-the-edge architecture, explicitly illustrating a refinement chain from \gls{hspl} to \gls{mspl} and then to low-level configurations, as part of a user-centric protection model.

Earlier work addressed policy-based management issues at a general level. Lupu et al.~\cite{lupu1999conflicts} have analyzed conflicts in policy-based distributed systems management, formalizing key classes of policy inconsistency and their detection. Closer to the network security domain, Basile et al.~\cite{basile2015conflictanalysis} have studied inter-technology conflicts in communication protection policies, revealing incompatibilities and redundancies across security enforced at different layers. On the verification side, Liu~\cite{liu2008formal} has proposed a formal approach to check whether firewall policies satisfy desired security properties. More broadly, Jabal et al.~\cite{jabal2019policyanalysis} have surveyed methods and tools for policy analysis, systematizing the research landscape with respect to correctness, consistency, and related analysis tasks.

In contrast to prior work centered on correctness, consistency, and policy analysis, this paper addresses the operational problem of automatically allocating and refining filtering policies over a heterogeneous network, jointly considering reachability constraints and device control support to derive minimal enforcement sets and device-specific configurations for deployment.

\cti is increasingly spread by means of dedicated sharing platforms, yet the resulting ecosystem remains non-uniform in scope, terminology, and offered capabilities, with limited empirical consolidation and standardization across vendor-driven solutions~\cite{sauerwein_threat_2017}. In parallel, the research community has proposed \cti frameworks and taxonomies to structure activities and resources, and to position widely used analytical models within broader processes~\cite{irfan_taxonomy_2022}. Despite this maturation, \cti workflows still face recurring challenges, ranging from data growth and diversity to the difficulty of converting unstructured intelligence into actionable defenses, highlighting the need for automation to bridge \cti production and concrete security operations~\cite{alguliyev_cti_2023}.

One important line of work indeed focuses on extracting structured knowledge from unstructured \cti reports. 
Rahman et al. have proposed ALERT, which targets the extraction of attack techniques from \cti narratives and their mapping to the MITRE ATT\&CK knowledge base; they have explicitly addressed the annotation issue by coupling \glspl{llm} with a classifier and active learning to reduce labeling effort~\cite{rahman_alert_2024}. 
Similarly, Zhang et al.'s EX-Action addresses the automatic extraction of threat actions from \cti reports, combining \gls{nlp} with a multimodal learning approach and introducing a notion of completeness for the extracted actions~\cite{zhang_ex-action_2021}. 
Complementary to technique/action-centric extraction, Marchiori et al. have presented STIXnet, which proposes a modular pipeline to extract heterogeneous \cti entities and relations; their solution aims at representing them as \gls{stix} objects, with the goal of achieving broad coverage of structured cyber threat knowledge beyond a small set of entity types~\cite{marchiori_stixnet_2023}. 
Along the same motivation, Jo et al. have implemented Vulcan, which studies automatic \cti extraction from unstructured text and argues that \gls{ioc}-centric extraction alone is often insufficient for threat understanding; they support the inclusion of richer \cti elements (e.g., attack vectors and tools) to enhance more informed downstream security processes~\cite{jo_vulcan_2022}.
While these approaches prove insightful, primarily advancing the upstream step of turning reports into structured intelligence (techniques, actions, or standardized \gls{stix} objects), the work discussed in this paper targets a different bottleneck: the passage from security intent into deployable network enforcement under operational constraints, which is not discussed in the aforementioned works.

%% file: sections/conclusions.tex
This paper presented our research on automated security policy refinement and allocation. The focus is on the burden of manual policy allocation in heterogeneous networks, where precise knowledge of network topology, devices, and available security controls is critical and often leads to misconfigurations.

The primary contribution is \refeng, a system that starts from high-level security intents and allocates proper security configurations across an input network model, while accounting for reachability constraints and device-specific control support. In the current instantiation, these artifacts are retrieved by an AI-based module of \refeng that extracts security-relevant entities from \cti reports. Two other modules are then invoked to (i) analyze the network topology and device capabilities, (ii) identify one or more minimal sets of devices that must be configured to enforce the policies, and (iii) generate medium-level policies to be enforced on the selected devices, accounting for the controls supported by those devices. Finally, one last module converts these medium-level policies, via a model-based process, into low-level policies for specific security controls, ready for deployment on the identified devices.

Overall, the proposed system reduces reliance on manual analysis and configuration by automatically deriving placements and control-specific policies from high-level intent. As a result, opportunities for human error and oversights of critical requirements are reduced.

The applicability of \refeng was validated through use cases involving packet and web filtering policies, which evaluated the allocation and enforcement of these policies across different network topologies. These policies were derived from \glspl{ioc} and threat descriptions in real \cti reports. The results support the correctness and practicality of the proposed workflow.

Future work will enhance the versatility of \refeng, thereby enabling the processing of additional classes of security policies. A deeper study of integrating the AI-based module for \cti processing and the model-based module for policy translation is also recommended and currently under consideration. 